\documentclass[fleqn,usenatbib]{mnras}

\usepackage{newtxtext,newtxmath}

\usepackage[T1]{fontenc}
\usepackage{ae,aecompl}

\usepackage{graphicx}	
\usepackage{amsmath}	
\usepackage{amssymb}	

\newcommand{\igr}{IGR J17062--6143}

\newcommand{\caps}[1]{{\scshape{#1}}}
\def\myr{M$_{\odot}$ yr$^{-1}$}

\title[Binary properties of IGR~J17062--6143]{Multiwavelength characterisation of the accreting millisecond X-ray pulsar and ultra-compact binary IGR~J17062--6143}

\author[Hern\'andez Santisteban et al.]{
J. V. Hern\'andez Santisteban,$^{1,2,3}$\thanks{E-mail: jvhs1@st-andrews.ac.uk (JVHS)}
V. C\'uneo,$^{4,5,6,7}$
N. Degenaar,$^{2,1}$
\newauthor
J. van den Eijnden,$^{1,2}$
D. Altamirano,$^{8}$
 M. N. G\'omez,$^{7,9}$
D M. Russell,$^{10}$
\newauthor
 R. Wijnands,$^{1}$ 
 R. Golovakova,$^{10}$
M. T. Reynolds$^{11}$ 
and J.M. Miller$^{11}$
\\
$^{1}$Anton Pannekoek Institute for Astronomy, University of Amsterdam, Science Park 904, NL-1098 XH Amsterdam, the Netherlands\\
$^{2}$Institute of Astronomy, University of Cambridge, Madingley Road, Cambridge CB3 OHA, UK\\
$^{3}$SUPA School of Physics \& Astronomy, University of St Andrews, North Haugh, St Andrews KY16 9SS, UK\\
$^{4}$Instituto de Astrof\'isica de Canarias (IAC), E-38205 La Laguna, Tenerife, Spain \\
$^{5}$Departamento de Astrof\'isica, Universidad de La Laguna, E-38206 La Laguna, Tenerife, Spain \\
$^{6}$Instituto Argentino de Radioastronom\'ia (CCT La Plata, CONICET), C.C.5, (1984) Villa Elisa, Buenos Aires, Argentina\\
$^{7}$Observatorio Astron\'omico de C\'ordoba, C\'ordoba, Argentina\\
$^{8}$Department of Physics and Astronomy, University of Southampton, Southampton, SO17 1BJ, UK\\
$^{9}$CONICET, Consejo Nacional de Investigaciones Cient\'ificas y T\'ecnicas, Argentina\\
$^{10}$New York University Abu Dhabi, PO Box 129188, Abu Dhabi, UAE\\
$^{11}$Department of Astronomy, University of Michigan, 1085 South University Avenue, Ann Arbor, MI  48109, USA
}

\date{Accepted 2019 July 15. Received 2019 July 8; in original form 2018 November 27}

\pubyear{2019}

\begin{document}
\label{firstpage}
\pagerange{\pageref{firstpage}--\pageref{lastpage}}
\maketitle

\begin{abstract}
\igr\ is an ultra-compact X-ray binary (UCXB) with an orbital period of 37.96 min. It harbours a millisecond X-ray pulsar that is spinning at 163 Hz and and has continuously been accreting from its companion star since 2006. Determining the composition of the accreted matter in UCXBs is of high interest for studies of binary evolution and thermonuclear burning on the surface of neutron stars. Here, we present a multi-wavelength study of \igr\ aimed to determine the detailed properties of its accretion disc and companion star. The multi-epoch photometric UV to near-infrared spectral energy distribution (SED) is consistent with an accretion disc $F_{\nu}\propto\nu^{1/3}$. The SED modelling of the accretion disc allowed us to estimate an outer disc radius of $R_{out}=2.2^{+0.9}_{-0.4} \times 10^{10}$ cm and a mass-transfer rate of $\dot{m}=1.8^{+1.8}_{-0.5}\times10^{-10}$ M$_{\odot}$ yr$^{-1}$. Comparing this with the estimated mass-accretion rate inferred from its X-ray emission suggests that $\gtrsim$90\% of the transferred mass is lost from the system. Moreover, our SED modelling shows that the thermal emission component seen in the X-ray spectrum is highly unlikely from the accretion disc and must therefore represent emission from the surface of the neutron star. Our low-resolution optical spectrum revealed a blue continuum and no emission lines, i.e. lacking H and He features. Based on the current data we cannot conclusively identify the nature of the companion star, but we make recommendations for future study that can distinguish between the different possible evolution histories of this X-ray binary. Finally, we demonstrate how multiwavelength observations can be effectively used to find more UCXBs among the LMXBs.

\end{abstract}



\section{Introduction}
\igr\ is a weak X-ray source that was discovered by \textit{Integral} in 2006. Only years later, it was identified as an accreting neutron star (NS) low-mass X-ray binary (LMXB) when a thermonuclear burst was detected with {\it Swift} in 2012 \citep{Degenaar:2012aa}. Since its discovery, the source seems to have been persistently accreting at a low luminosity of $\sim10^{-3}$ $L_{\rm Edd}$  \citep{Remillard:2008aa,Degenaar:2012aa,Degenaar:2016aa}, and two energetic, long thermonuclear X-ray bursts have been detected \citep[in 2012 and 2015;][]{Degenaar:2013aa,Keek:2016aa}. 

A detailed study of the X-ray spectrum of \igr\@ revealed a broad Fe-K emission line near $\simeq6.5$ keV. Modeling this feature as X-rays reflecting off the accretion disc \citep{Fabian:2010aa}, suggests that the inner edge of the disc is truncated at $R_{in}\gtrsim225$ km ($\gtrsim100 R_g$) from the neutron star \citep[][]{Degenaar:2016aa,eijnden:2017aa}. This contrasts with LMXBs accreting at higher rates, where the inner disc typically resides a factor $\gtrsim 5$ closer to the compact primary \citep[e.g.][for sample studies]{cackett2010,ludlam2017}. The truncation is likely a direct effect of the low accretion rate, e.g. resulting from the formation of a radiatively-inefficient accretion flow. Alternatively, the magnetosphere of the neutron star may push the accretion disc out \citep{Degenaar:2016aa}. At the low accretion rate inferred for \igr, this would require a magnetic field strength of $B\geq2.5\pm2.1\times10^8$ G \citep{eijnden:2017aa}.

A single archival 1.2 ks \textit{RXTE} observation revealed coherent X-ray pulsations at a frequency of 163.65 Hz \citep{Strohmayer:2017aa}, suggesting that \igr\@ is an accreting millisecond X-ray pulsar (AMXP). The AMXPs form a sub-class of LMXBs in which the neutron star is able to channel plasma from the accretion disc along its magnetic field lines onto its magnetic poles \citep[e.g.][]{Wijnands:1998aa}. The presence of X-ray pulsations was confirmed with new data from \textit{NICER}, and suggests that the neutron star has a magnetic field strength of $B\leq3.8\times10^8$ G \citep{Strohmayer:2018aa}. This is similar to that of other AMXPs \citep[e.g. SAX~J1808-359;][]{Mukherjee:2015aa}, and consistent with the inner disc radius measurement from X-ray reflection modeling \citep{Degenaar:2016aa,eijnden:2017aa}. 

Timing of the X-ray pulsations allowed to measure an orbital period of 37.96 min \citep{Strohmayer:2018aa}. LMXBs with such short orbital periods are referred to as ultra-compact X-ray binaries (UCXBs) and can only harbour H-poor donor stars \citep[e.g.][]{Nelson:1986aa,Nelemans:2010aa}. Based on its sustained low accretion rate, which is easier to achieve if the accretion disc is small \citep[e.g.][]{Tsugawa:1997aa,intzand2007,Hameury:2016aa}, and the enhanced oxygen abundance inferred from high-resolution X-ray spectral modeling, it was previously hypothesized that \igr\@ could be an UCXB \citep{eijnden:2017aa}. UCXBs are a particularly intersting sub-population of LMXBs that are expected to be promising targets for future gravitational wave interferometry experiments \citep[e.g.][]{nelemans2003}. Furthermore, UCXBs are interesting laboratories to study the ashes of stellar nuclear burning \citep[e.g.][]{Deloye:2003aa}, and binary evolution models \citep[e.g.][]{Nelemans:2010mnras}, as well as models of thermonuclear burning on the surface of neutron stars \citep[e.g.][]{cumming2003,cumming2006}. 

In this work, we present a multi-wavelength study of \igr. In Section~\ref{sec:observations}, we present observations from X-ray to near-IR (NIR). In Section~\ref{sec:analysis} we construct an average SED and present our optical spectrum, which we use to constrain the properties of the accretion disc and donor star. In Section~\ref{sec:discussion}, we discuss the implications for the evolutionary history of \igr\@, and for thermonuclear burning models, and demonstrate that multi-wavelength observing campaigns can effectively be used to find new candidate UCXBs among the LMXB population.    

\section{Observations}\label{sec:observations}
\subsection{Faulkes Optical photometry}
We observed the field of \igr\ with the 2-m robotic Faulkes Telescope South (FTS), at Siding Spring, Australia, on 2016 October 4 and 2016 October 6. On both dates, 300-sec exposures were made in three filters; Sloan Digital Sky Survey (SDSS) $g^{\prime}$, $r^{\prime}$ and $i^{\prime}$-bands. FTS was equipped with a camera with a pixel scale of 0.304 arcsec pixel$^{-1}$ and a field of view of $10\times 10$ arcmin. The images were de-biased and flat-fielded using the automatic Las Cumbres Observatory (LCO) pipeline {\sc banzai}. The seeing as measured from the images was 2.8$^{\prime\prime}$ and 1.9$^{\prime\prime}$ on 4 and 6 October, respectively.

We detected a faint source consistent with the position of \igr\@ in all images. The spatial resolution in the FTS images is too poor to separate out the faint nearby source detected in our Magellan NIR images, but this object likely has a negligible contribution to our optical images (see Section~\ref{sec:magellan}). The optical counterpart is bluer than the surrounding field stars (shown in the bottom panel of Fig.~\ref{fig:finding_chart}), which is consistent with emission from an accretion disc. Photometry was carried out using \small PHOT \normalsize in \small IRAF\normalsize\footnote{\href{http://iraf.noao.edu/}{http://iraf.noao.edu/}}. Flux calibration was achieved using the known $g^{\prime}$, $r^{\prime}$ and $i^{\prime}$ magnitudes of four stars in the field of view tabulated in the AAVSO Photometric All-Sky Survey \citep[APASS;][]{APASS}. The resulting magnitudes are given in Table \ref{tab:flux_table}. The errors include the 1$\sigma$ uncertainty in the comparison star magnitudes, which are small; the error is dominated by the S/N of the X-ray binary.

\subsection{Swift UV photometry}\label{sec:uvot}
\igr\ has been observed multiple times with \textit{Swift} since its discovery in 2006, which provides a rich photometric optical and UV characterisation of the system. We downloaded all the available calibrated UVOT data files from the \textit{Swift Data Centre} and performed aperture photometry using \caps{uvotsource} (as implemented in \caps{HEASOFT} v6.18)  using a 4 arcsec aperture for the target source and an 11 arcsec aperture for the background region. Although the full data set includes measurements during two type-I X-ray bursts \citep{Degenaar:2013aa,Keek:2016aa}, we have excluded them for the analysis in the following sections. We present the average fluxes in every UVOT filter in Table~\ref{tab:flux_table}, where the quoted errors reflect the r.m.s. The individual measurements are shown in Table~\ref{tab:flux_swift}.

\subsection{Magellan NIR photometry}\label{sec:magellan}
We obtained near-infrared (NIR) photometry with the FourStar camera \citep{Persson:2013aa} at the 6.5m Baade \textit{Magellan} Telescope in Cerro las Campanas, Chile. The images were taken in two observing campaigns, 2013 June 16 and 2014 May 8. We observed the source in three filters $J$, $H$ and $Ks$; the details of the individual observations are presented in Table \ref{tab:flux_table}. The telescope was nodded in a AB-AB mode, in order to optimise sky subtraction. The \caps{iraf}/\caps{fsred} package (provided by Andy Monson) was used to de-bias, flat-field, align, and co-add the FourStar observations for each object and filter. Aperture photometry was performed using 2MASS sources to determine the zero-point. 

\begin{figure}
 \centering
	\includegraphics[trim=0.1cm 0.0cm 0.0cm 0.0cm, clip,width=\columnwidth]{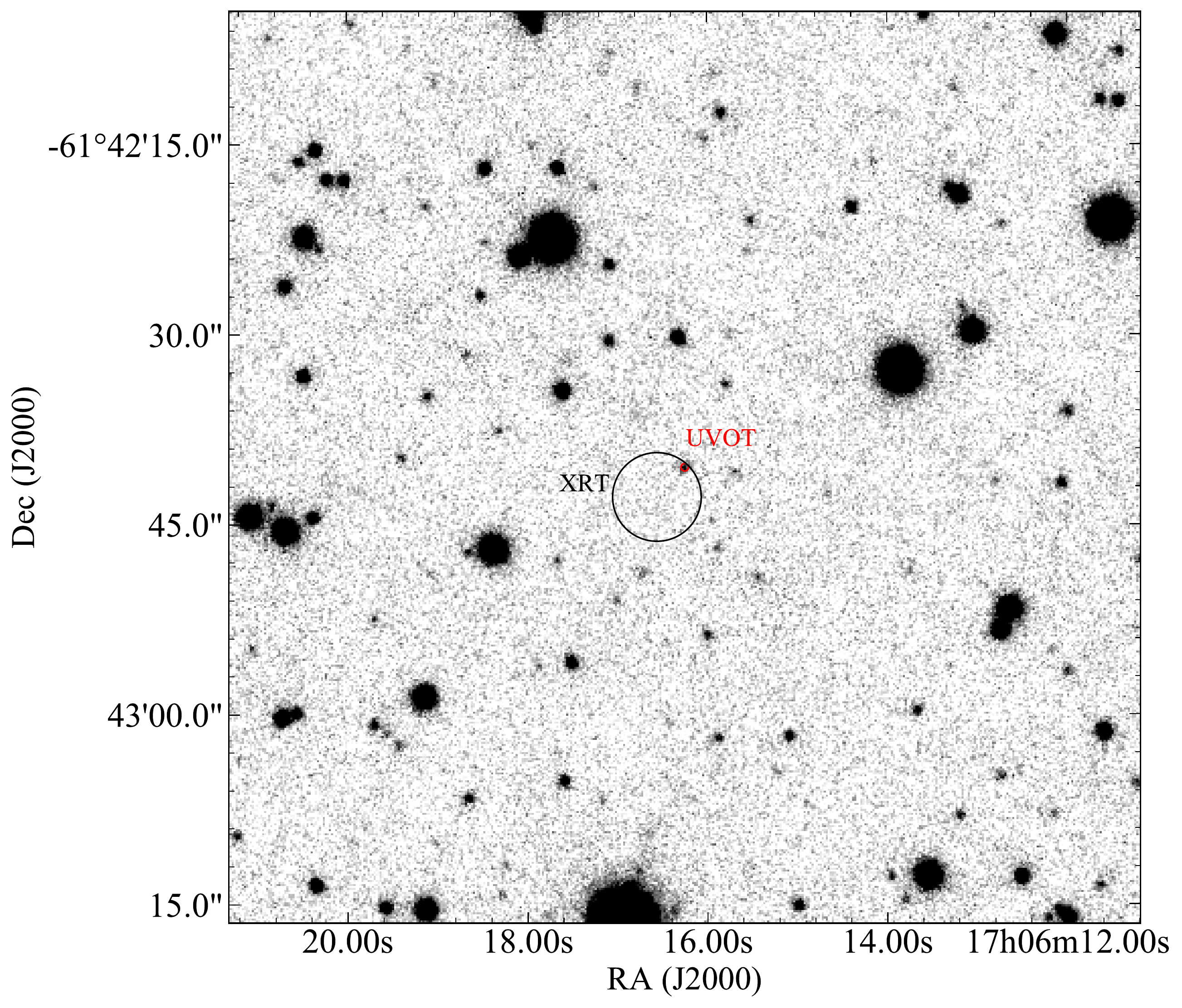}
    \includegraphics[trim=0.1cm 0.0cm 0.0cm 0.0cm, clip,width=\columnwidth]{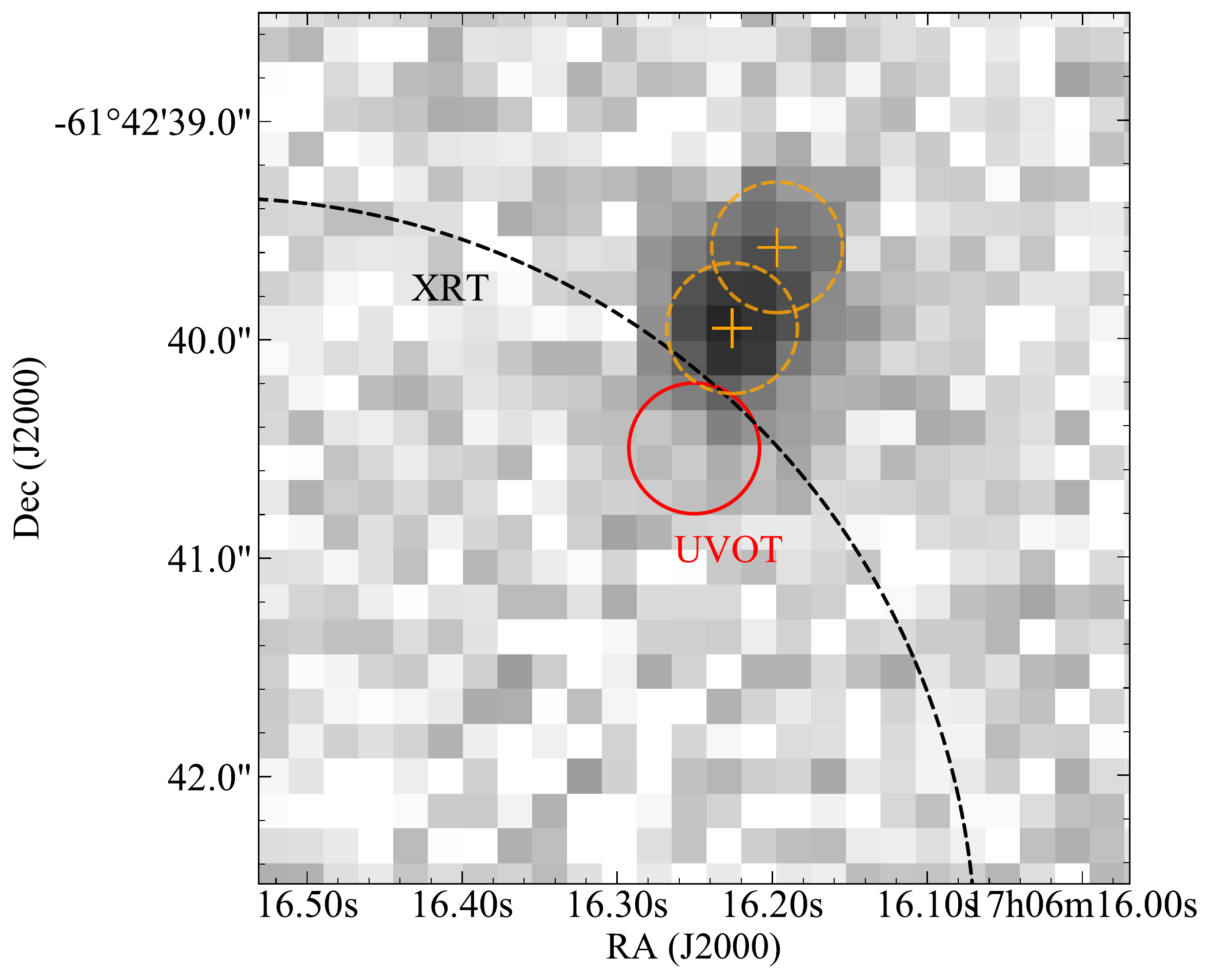}
    \includegraphics[trim=0.1cm 0.0cm 0.0cm 0.0cm, clip,width=\columnwidth]{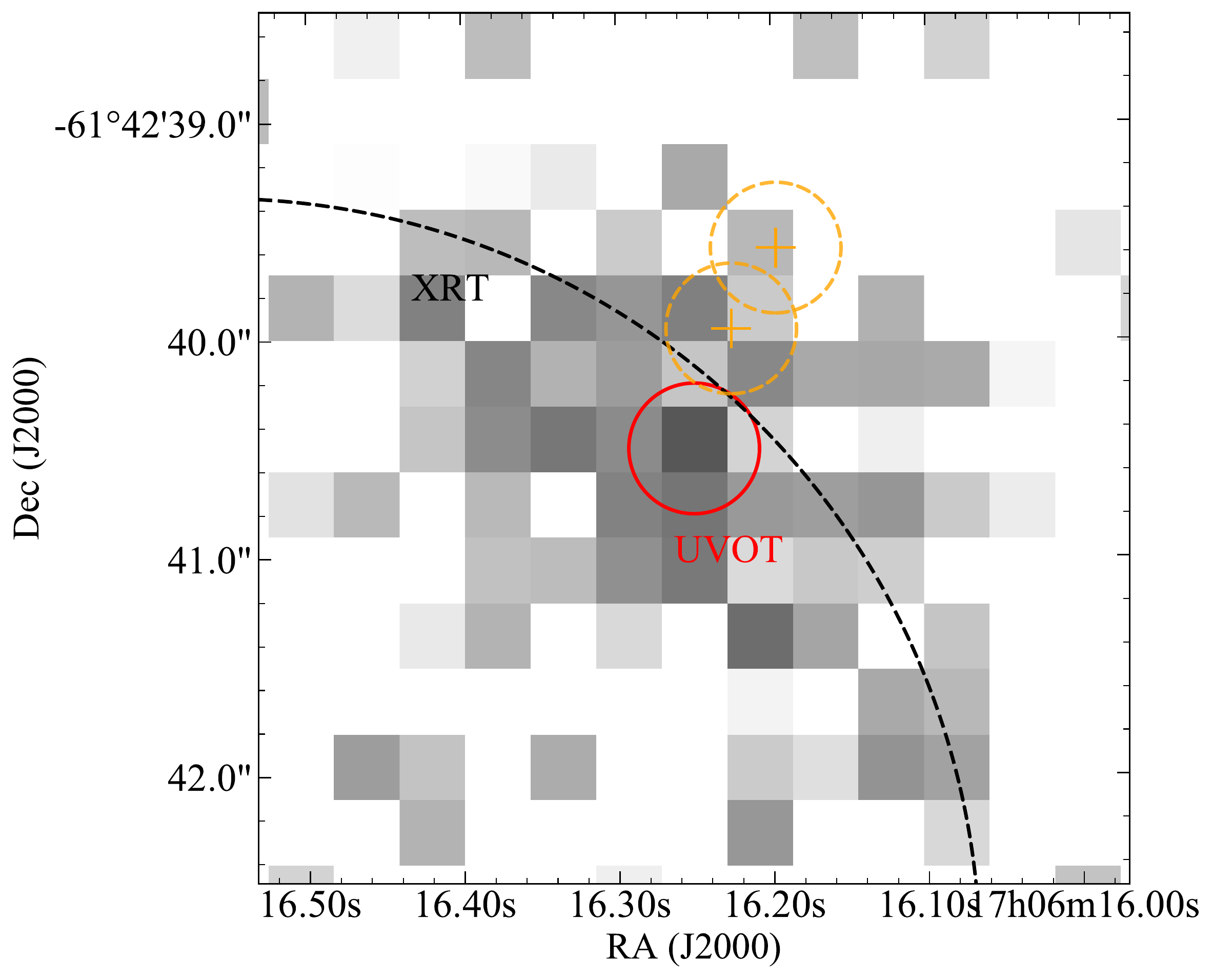}
     \caption{Finding charts for \igr\@. \textit{Top:} Wide field of view $96\times96$ arcmin with Magellan in the $J$ filter. \textit{Middle:} Zoom of the top panel on the position of \igr\ and the blended source to the north-west. The circles on these sources represent the 0.3" uncertainty from the overall astrometric solution. We show the confidence regions as  determined by \textit{Swift}/UVOT (red) and \textit{Swift}/XRT (black) is also shown for reference \citep{Ricci:2008aa}. \textit{Bottom:} Optical Faulkes photometry ($r'$ band). The source is compatible with \igr\ position obtained from the Magellan images marked as red circles.}
    \label{fig:finding_chart}
\end{figure}

\begin{figure}
    \includegraphics[trim=0.1cm 3.6cm 0.0cm 3.5cm, clip,width=\columnwidth]{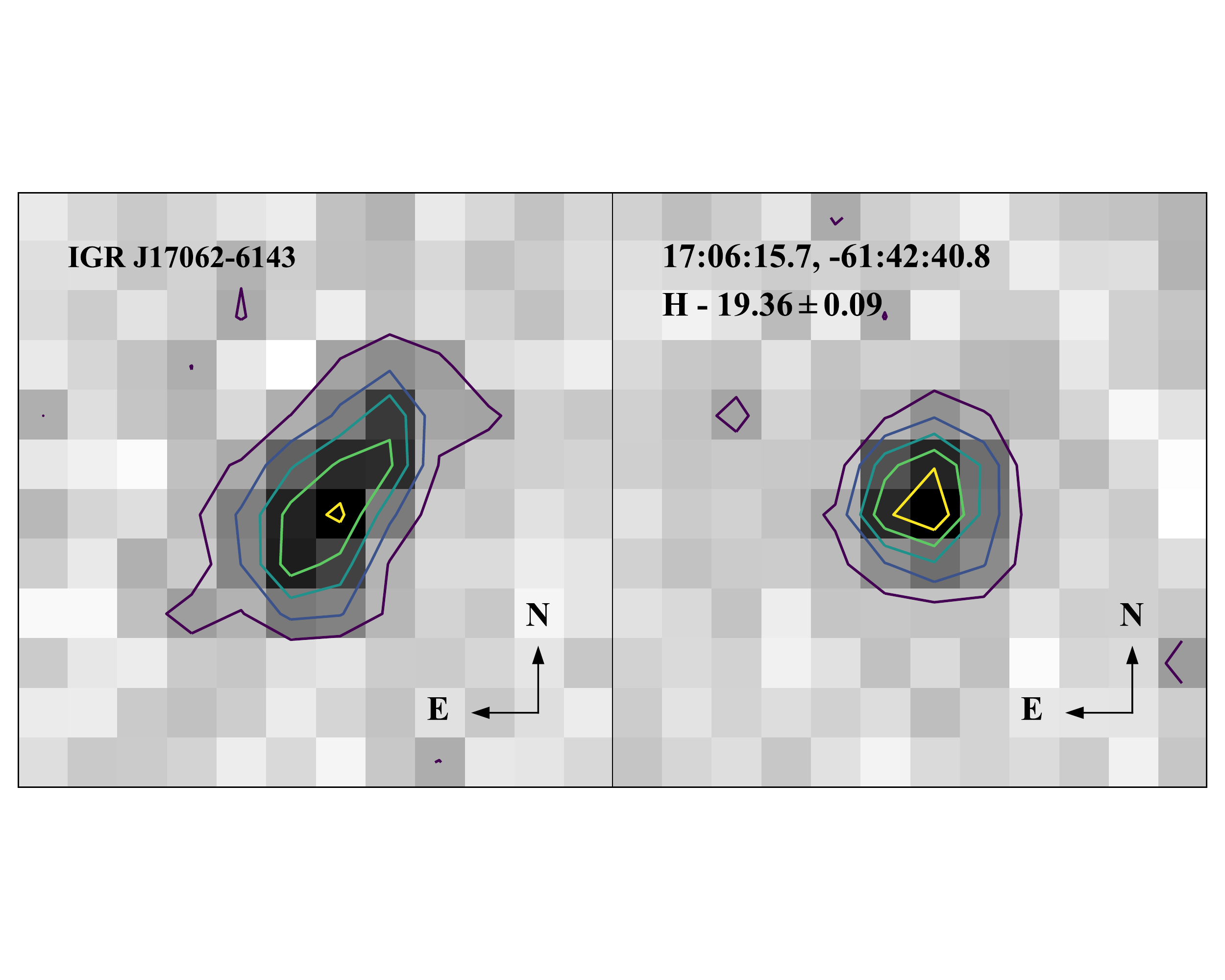}
    \caption{NIR source identification of \igr\ using Magellan. \textit{Left:} $K$-band imaging revealing a fainter blended source to the north-west. \textit{Right:} A close-up to a star with similar brightness which shows the average PSF of the image. Contours track 10, 20, 30 and 40 counts above the background for both panels.
}
    \label{fig:finding_chartclose}
\end{figure}

We clearly detect a source at the position of \igr\ in all three filters, as shown in the NIR finding chart in Fig.~\ref{fig:finding_chart}. 
The high quality of the NIR images, with a measured seeing of $\sim0.5$ arcsec, and a smaller pixel scale (0.159" pixel$^{-1}$)~revealed a fainter source blended in the north-west direction as shown in Fig.~\ref{fig:finding_chartclose}. We performed PSF photometry using \caps{daophot} to extract the individual measurements, given in Table~\ref{tab:flux_table}. The centroid for brightest, Southern source is $\alpha=$17:06:16.226(16), $\delta=$-61:42:39.95(23) and for the fainter Northern source is $\alpha=$17:06:16.197(22), $\delta=$-61:42:39.58(29). The errors on the positions represent the 90\% upper bound statistical uncertainty on the centroid for all stars detected at a similar magnitude. We also note a $\sim0.3$" rms uncertainty for astrometric solution over the entire field. The latter are explicitly shown in the middle and bottom panels of Fig.~\ref{fig:finding_chart}. 

The brighter NIR source is more consistent with the position of the blue optical counterpart detected in our Faulkes images, and also with its \textit{Swift/UVOT} coordinates. We therefore identify the brighter NIR source as the counterpart to \igr. The fainter source has 2MASS colours $J-H\,\sim\,0.9$ and $H\,-\,Ks\,\sim\,0.45$ (magnitudes in Table~\ref{tab:flux_table} are shown in AB system)\footnote{In the 2MASS photometric system, this fainter source $J=21.95\pm0.17$, $H=21.05\pm0.42$ and $K_s=20.55\pm0.39$ }, consistent with a late-M star. Our photometric method minimises the contribution in the NIR and given the colours of this source, the contribution in the optical and UV bands is negligible.

\begin{table*}
	\centering
	\caption{Observation log of the multi-wavelength photometry of \igr. We present only the average magnitudes and r.m.s. (quoted as the error) for the \textit{Swift}/UVOT data. Otherwise, all errors represent the 1$\sigma$ confidence level. For a complete breakdown of the \textit{Swift/UVOT} observations, see Table~\ref{tab:flux_swift} in the Appendix. }
	\label{tab:flux_table}
\begin{tabular}{lccccccll}
\hline
Facility & Filter & $\lambda_0$ & Date       & Exposure &  		Magnitude 		 &  &  &  \\
         &        & \AA         & UTC        & s        & AB mag             &  &  &  \\
\hline
Swift    & UVW2   & 1928        &            &          & 20.10 $\pm$ 0.05   &  &  &  \\
         & UVM2   & 2246        &    see     &      see & 20.40 $\pm$ 0.15   &  &  &  \\
         & UVW1   & 2600        &Appendix    &  Appendix& 20.03 $\pm$ 0.08 &  &  &  \\
         & U      & 3465        &\S~\ref{sec:A2}&\S~\ref{sec:A2}& 19.85 $\pm$ 0.39&  &  &  \\
         & B      & 4349        &            &          & 20.30 $\pm$ 1.69   &  &  &  \\
\hline
Faulkes  & g'     & 4770        & 2016-10-04 &  300     & 20.23 $\pm$ 0.16   &  &  &  \\
         &        &             & 2016-10-06 &  300     & 20.29 $\pm$ 0.13   &  &  &  \\
         & r'     & 6231        & 2016-10-04 &  300     & 20.30 $\pm$ 0.17   &  &  &  \\
         &        &             & 2016-10-06 &  300     & 20.12 $\pm$ 0.10   &  &  &  \\
         & i'     & 7625        & 2016-10-04 &  300     & 20.37 $\pm$ 0.19   &  &  &  \\
         &        &             & 2016-10-06 &  300     & 20.05 $\pm$ 0.12   &  &  &  \\
\hline
Magellan & J      & 12350       & 2013-06-16 & 87      & 20.54 $\pm$ 0.08   &  &  &  \\
   \igr  &        &             & 2014-05-17 & 306      & 20.84 $\pm$ 0.04   &  &  &  \\         & H      & 16620       & 2013-06-16 & 306      & 20.63 $\pm$ 0.08   &  &  &  \\
         &        &             & 2014-05-17 & 157      & 20.44 $\pm$ 0.05   &  &  &  \\
         & Ks     & 21590       & 2013-06-16 & 218      & 20.98 $\pm$ 0.11   &  &  &  \\
         &        &             & 2014-05-17 & 157      & 20.78 $\pm$ 0.10   &  &  &  \\
\hline
Magellan & J      & 12350       & 2014-05-17 & 87             & 23.39 $\pm$ 0.18   &  &  &  \\
Fainter source         & H      & 16620       & 2014-05-17 & 306      & 22.96 $\pm$ 0.44   &  &  &  \\
         & Ks     & 21590       & 2014-05-17 & 218      & 22.93 $\pm$ 0.39   &  &  &  \\

\hline
	\end{tabular}
\end{table*}

\subsection{Swift X-ray spectroscopy}\label{sec:xrt}
We extracted all available \textit{Swift}/XRT spectra of \igr\@ between the NIR and optical observations to obtain a long-term X-ray light curve  of the source. We used the \textit{Swift}/XRT Online Data Products Generator\footnote{See http://www.swift.ac.uk/user\_objects/index.php} \citep{Evans:2009aa} to extract $19$ WT-mode and $23$ PC-mode spectra from in total $32$ epochs. To each spectrum, we fitted a simple absorbed blackbody plus power law model [\textsc{tbabs*(bbodyrad+powerlaw)}] in \textsc{xspec} v12.9.0 \citep{arnaud96} and calculated the unabsorbed flux in the $2.0$--$10.0$ keV range. We assumed an absorbing hydrogen column density of $N_H = 2.3\times10^{21}$ $\rm cm^{-2}$ \citep{Degenaar:2016aa}. 

\begin{figure}
	\includegraphics[width=\columnwidth]{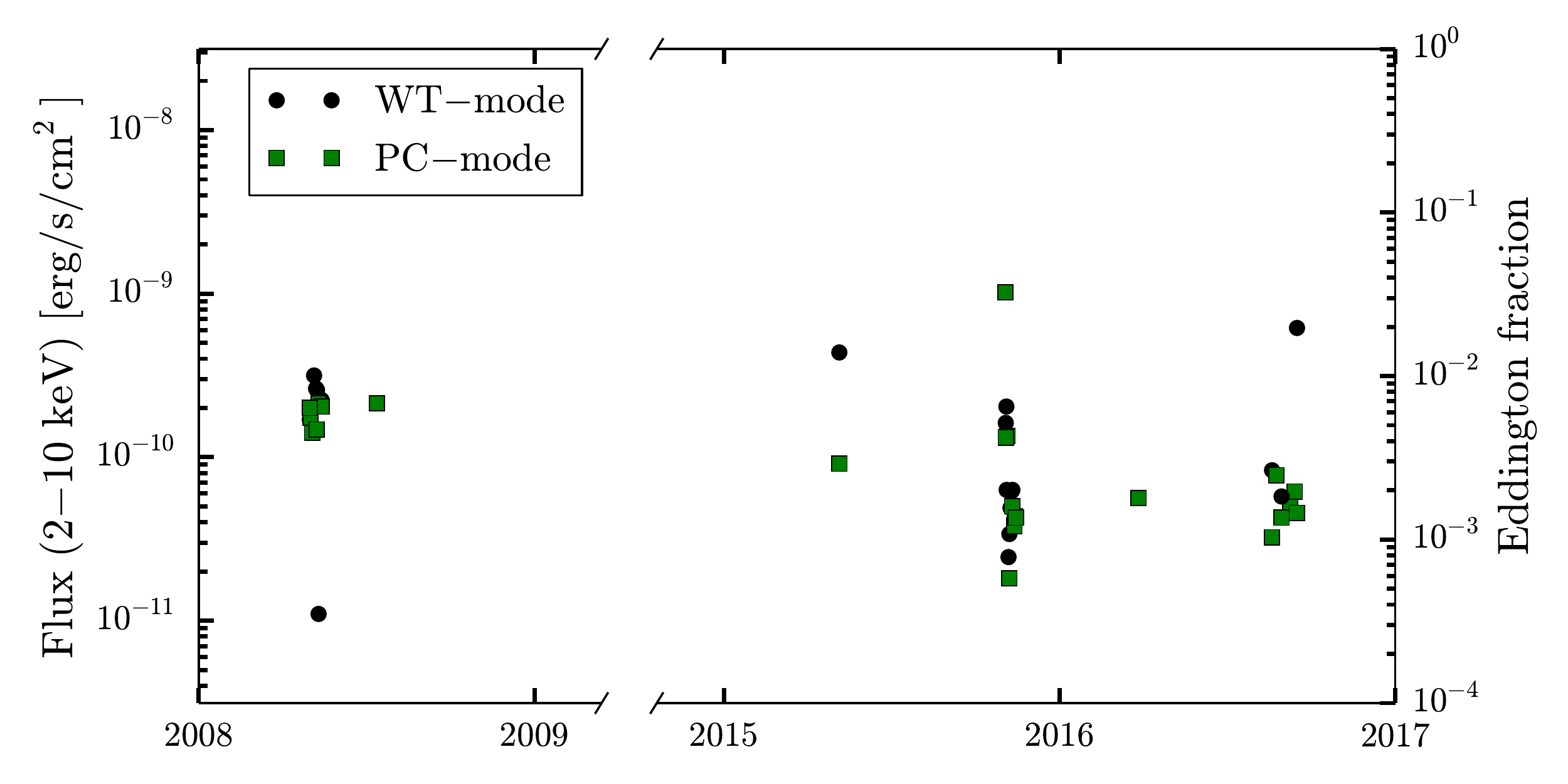}
    \caption{Long-term light curve of \igr\@ as observed with $Swift$/XRT. Note that for visual clarity, we exclude the single observation during the early decay stages of the source's 2012 thermonuclear X-ray burst \citep[reported by][]{Degenaar:2016aa}. The observations at the end of 2015 were taken during the late decay a second thermonuclear X-ray burst citep[reported by][]{Keek:2016aa}; these data are shown in this plot but not used in the analysis in Section~\ref{sec:analysis}.}
    \label{fig:xray_lc}
\end{figure}

In Figure \ref{fig:xray_lc} we show the long-term light curve, where the Eddington ratio was calculated assuming the empirical Eddington luminosity of $L_{\rm Edd} = 3.8\times10^{38}$ erg s$^{-1}$ \citep{kuulkers2003}, a distance of $7.3$ kpc \citep{Keek:2016aa}, and a bolometric correction factor of 2 to convert from $2 - 10$ keV to bolometric fluxes \citep[e.g.][]{intzand2007,galloway2008}. 
It can be seen that the X-ray luminosity varied in a range of $\sim10^{-3} - 10^{-2}$ $L_{\rm Edd}$ over the past decade. 

\subsection{Gemini optical spectroscopy}\label{sec:opt_spec}
We obtained long-slit spectroscopic observations of \igr\@ with the Gemini Multi-Object Spectrograph (GMOS) at the 8m Gemini South telescope under a Fast Turnaround program (GS-2016A-FT-24) on 2016 September 27. The GMOS instrument uses the Hamamatsu CCD ($6266\times4176$ pixels). Six spectra of 900s, three centred at 570 nm and three at 580 nm to avoid the chip gaps, were taken using the the B150 grating (150 l/mm), a slit width of $1''$ and a $2\times2$ binning. The GG455\_G0329 blocking filter was also used in order to avoid second order overlap. The chosen set-up resulted in the spectral coverage of the $4400-10800$ \AA\ wavelength range and a spectral resolution of 3 \AA. The seeing during the observation was  $\sim0.5-0.6''$. A CuAr lamp was also observed for each configuration in order to perform the wavelength calibration.

Spectra were reduced using the \caps{iraf}-\caps{gemini} package. 
The flux calibration of each spectrum was executed using observations of a standard star, also taken as part of the program and the errors propagated through the \caps{gemini} pipeline. Finally, in order to increase the signal-to-noise (S/N), the six spectra were combined to obtain a final spectrum with a S/N of $\sim60$ at 6500 \AA\ and 7500 \AA, and $\sim40$ at 8500 \AA. The final spectrum was cut to keep the spectral region between 5000 and 9000 \AA, as shown in Fig.~\ref{fig:optical_spec}. The bluest part of the spectrum was removed as the filter lowered the response at those wavelengths. On the other side, the reddest part was cut out because the atmospheric absorption becomes too significant and distorts the slope of the continuum. A few features were left in the final spectrum, marked as crosses  in  Fig.~\ref{fig:optical_spec}. We checked the individual spectra before combining in order to asses the validity of every singular feature. Most of them were tracked to bad background subtraction and cosmic ray removal. We only find the \ion{Na}{i} doublet $\lambda$5889, $\lambda$5895 \AA\ to be present in all six spectra. Given the intrinsic variability of the object and non-simultaneity of the observations (see Section~\ref{sec:sed}), we did not attempt to use the Faulkes photometry to perform a correction on the flux calibration.

\section{Results}\label{sec:analysis}
\subsection{A featureless optical spectrum}
\begin{figure*}
    \includegraphics[trim=0.0cm 0.0cm 0.0cm 0.1cm, clip,width=15cm]{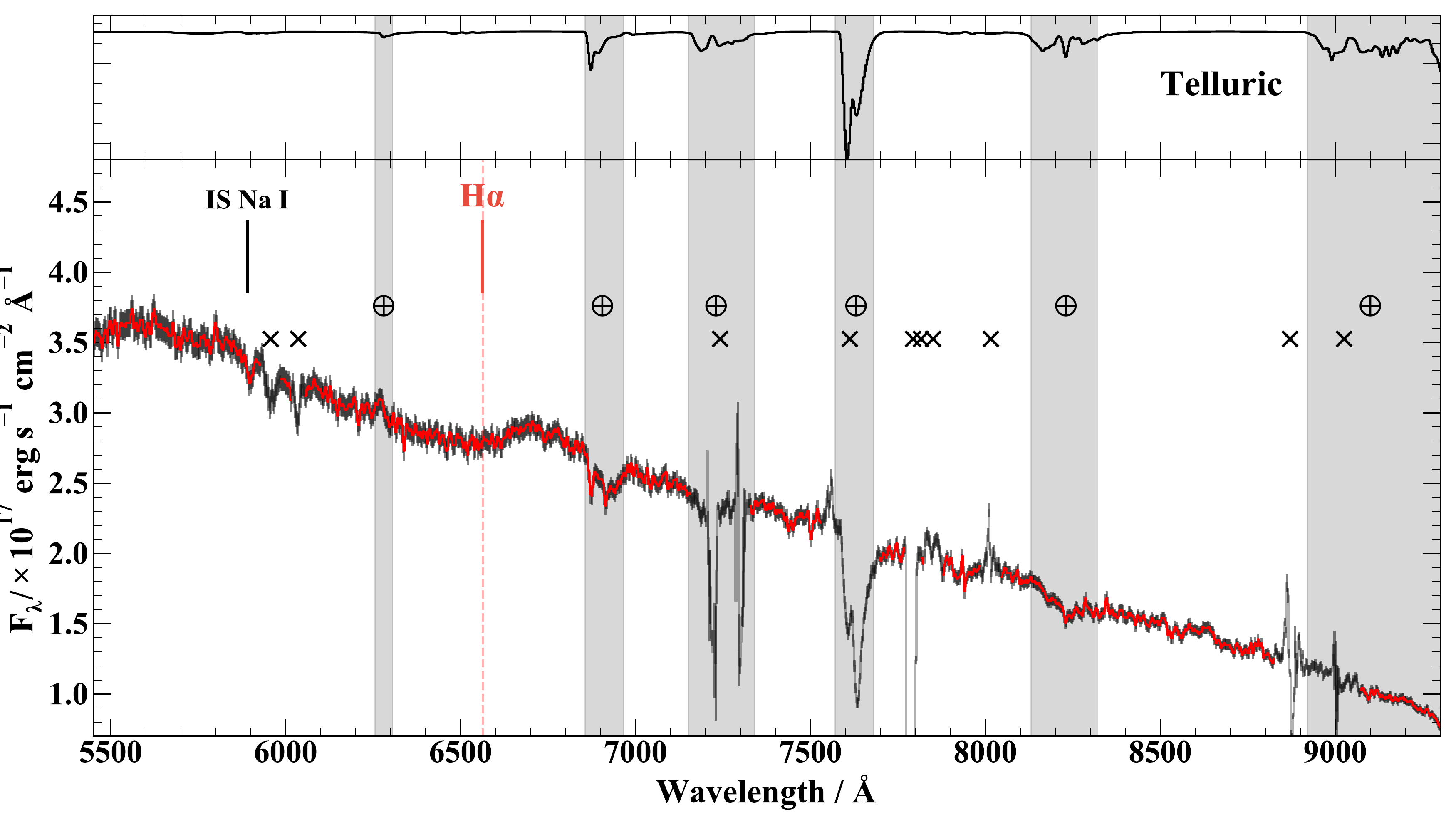}
    \caption{
    Flux calibrated optical spectrum of \igr. $Top:$ Telluric transmission spectrum is shown for reference. $Bottom:$  Regions of telluric absorption have been labelled with $_{\bigoplus}$  and shaded in grey. The crosses show instrumental artefact and bad sky-line subtraction features  (see text for details). The locations of H$\alpha$ and interstellar \ion{Na}{i} are marked as well.}
    \label{fig:optical_spec}
\end{figure*}

The optical spectrum of \igr, displayed in Fig.~\ref{fig:optical_spec}, shows a blue continuum with no emission line features. We only identify one absorption feature, due to interstellar extinction of \ion{Na}{i} doublet $\lambda$5889, $\lambda$5895 \AA\@. We find a feature around 7816~\AA, which is coincident with a known \ion{He}{i} line. However, after looking at the individual data (only three of the six had information in this region since it lies on the detector gap for the others), we find that the feature is an instrumental artefact. 

The optical spectra of LMXBs are typically dominated by that of the irradiated accretion disc, showing a blue continuum with strong Balmer, He{\sc ii} and Bowen emission features \citep[see e.g.][]{charles2006}. Occasionally, LMXBs can show broad absorption features when actively accreting \citep[see e.g.][]{Cornelisse:2009aa}. However, studies of  confirmed and candidate UCXBs have revealed similarly featureless optical spectra as we see for \igr\ \citep[e.g.][]{nelemans2004,Nelemans:2006aa}. Whereas the optical spectra of white dwarf analogues of UCXBs (so-called AM CVns) show rich line spectra, detecting emission lines of He, C, N, and O, in optical spectra of UCXBs appears to be only achievable when high signal to noise data is available \citep[e.g.][]{nelemans2004}. 

\subsection{Spectral energy distribution}\label{sec:sed}
\begin{figure*}
	\includegraphics[trim=0.5cm 0.2cm 0.0cm 0.2cm, clip,width=18cm]{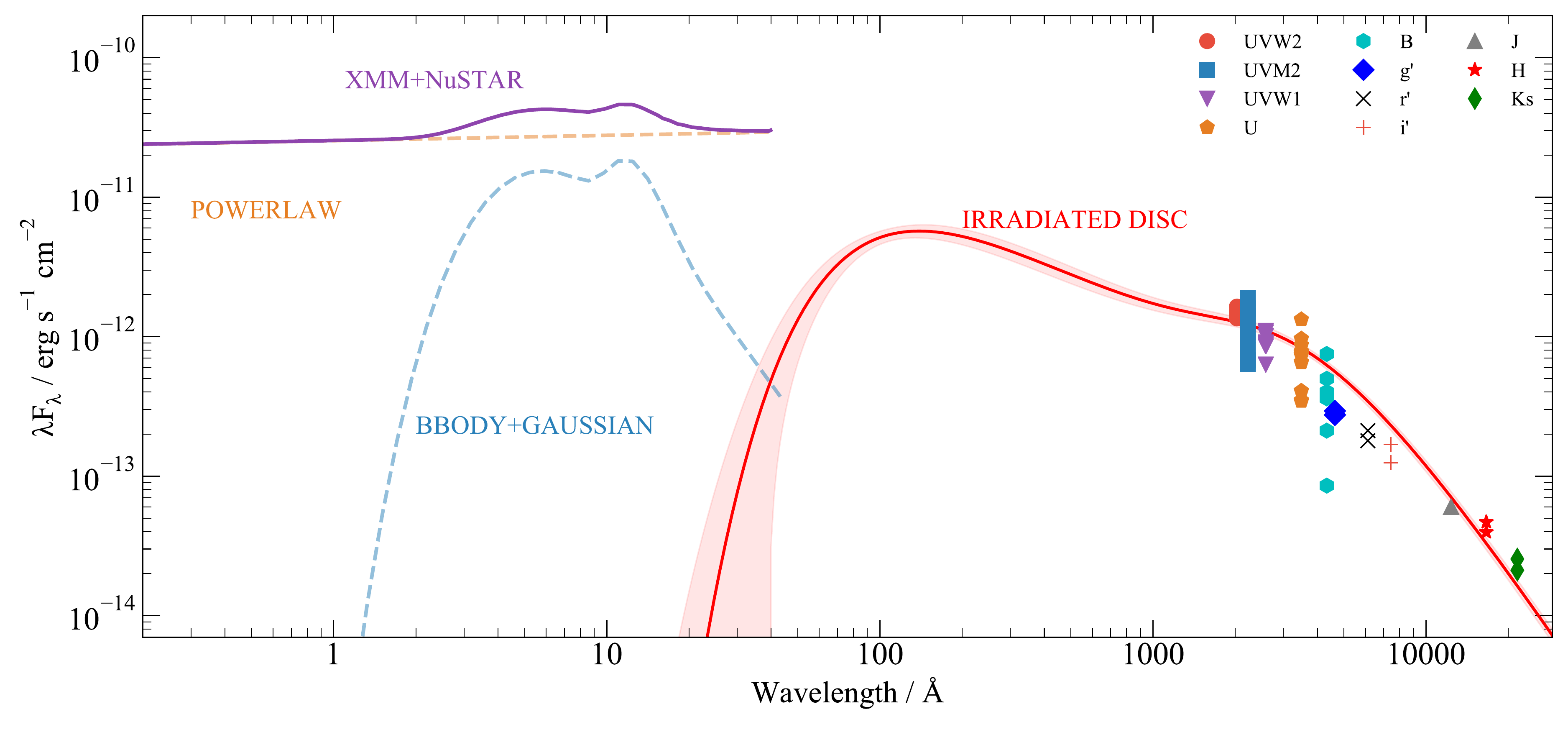}
    \caption{Broadband SED fitting of \igr\@ modelled as an irradiated accretion disc shown as the best fit model (red line). The UV to NIR data has been dereddened by E(B-V)=$0.135$. The X-ray spectrum was simulated from the best fit to simultaneous \textit{XMM-Newton} and \textit{NuStar} data (\caps{powerlaw+bbody+gaussian}) taken from \citet{eijnden:2017aa}. The red shaded region shows random realisations from our MCMC analysis. 
    }
    \label{fig:SED_fit}
\end{figure*}

We used our multi-wavelength data to construct the SED of \igr. 
The broadband SED of \igr\@ is shown in Fig.~\ref{fig:SED_fit}. We fit the UV to NIR data to a phenomenological extincted  power-law model of the form $F_{\lambda}\propto \lambda^{\Gamma}$. We used $\chi^2$ as our goodness-of-fit parameter and the 1$\sigma$ confidence intervals were obtained by scaling the errors so $\chi^2_{\nu}=1$. 
Based on the hydrogen column density inferred from analysing X-ray spectral data, $N_H = 2.3\times10^{21}$ $\rm cm^{-2}$ \citep{Degenaar:2016aa} and consistent with the line of sight extinction, we apply a reddening of E(B-V)$=0.135$ \citep{willingale:2013}. We then obtain a power-law index $\Gamma=-2.6\pm0.2$, consistent with the plateau formed by a steady-state accretion disc $F_{\lambda}\propto\lambda^{-7/3}$ \citep{Lynden-Bell:1969aa,Frank:2002aa}. Moreover, the surface temperature of the donors in UCXBs is expected to be $<10^4$~K after they evolve past their period minimum, and then quickly cool to just a few $10^3$~K, for periods $>20$~min \citep[][]{deloye2007}. In combination with their small radius ($\sim0.040~R_{\odot}$), we expect their emission to contribute $\lesssim$1\% in all the UV--NIR range. Combined with the index obtained from our power-law fit to the SED, we therefore assume that the UV-NIR wavelength range is dominated entirely by the accretion disc.

In order to retrieve physical parameters, which are more informative than a simple power law, we have used the model of an irradiated accretion disc as in \citet[][Eq. 10 to 15]{Chakrabarty:1998aa}. This model consists of an optically thick, geometrically thin accretion disc \citep{Shakura:1973lr,Frank:2002aa}, which emission is modified by the central X-ray source. Therefore, the temperature profile is defined by the combination of internal viscous heating and shallow X-ray heating. We have assumed a disc albedo $\nu_d=0.25$ and a canonical mass for the NS of M$_{NS}=1.4$ M$_{\odot}$. In order to compare the disc model to the available photometry ($m_n$ and $\sigma_n$, their associated 1$\sigma$ uncertainties), we applied the reddening of the object to the model spectrum and performed synthetic photometry for every filter in our sample. We have excluded the $Gemini$ spectrum from our analysis due to an unreliable absolute flux calibration (see Sec.~\ref{sec:opt_spec}). 

To derive the best fit parameters, we explored the parameter space by using a MCMC procedure as implemented in \caps{emcee} \citep{Foreman-Mackey:2013aa}\footnote{\url{http://dan.iel.fm/emcee/current/}.}. 
The likelihood function, $\mathcal{L}$, that we employed to retrieve the best fit parameters, where the uncertainties are Gaussian and independent, is given by
\begin{equation}
\ln\mathcal{L} = -\frac{1}{2}  \left[\sum_{n=0}^N \frac{(D_{n} - m_n)^2 }{\sigma_{T,n}^2} + \ln\left(2\pi\sigma_{T,n}^2 \right) \right]\,.
\end{equation}
where model prediction for each filter is $D_{n}$ and the total variance is defined as  $\sigma_{T,n}^2 = \sigma_n^2 + f^2m_{n}^2$. We have added a fractional scatter, $f$, common to the dataset, in order to reflect any intrinsic variability and non-simultaneity of the observations. Also, this added scatter will reflect any contamination arising from the dim star detected at NIR wavelengths (see \S\ref{sec:magellan}). 
The choice of priors for our MCMC procedure is described below.

During the energetic 2015 thermonuclear burst, signatures of photospheric radius expansion were observed, which is an indication that the Eddington limit was reached and puts the system at a distance of $7.3\pm0.5$ kpc \citep{Keek:2016aa}. 
This measurement is likely more accurate than the $\sim5$ kpc estimate inferred from the 2012 thermonuclear burst \citep{Degenaar:2013aa}, because the 2015 burst had softer energy coverage near its peak (using \textit{MAXI}; 2--20 keV) than the 2012 one (using \textit{Swift}/BAT; 15--50 keV), and should thus provide more reliable constraints on the soft $\sim2-3$~keV black body emission. We therefore used the \citet{Keek:2016aa} measurement and its associated $1\sigma$ error as a Gaussian prior for the distance. We again use E(B-V)$=$0.135 based on X-ray spectral fitting results \citep{Degenaar:2016aa} and assuming $R_V=3.1$ \citep[typical for the Milky Way;][]{Predehl:1995aa}. Furthermore, we used an inner disk radius of $R_{in}$, based on X-ray reflection analysis \citep[][]{Degenaar:2016aa,eijnden:2017aa}.\footnote{When left as a free fit parameter in our SED modelling, the inner disc radius was poorly constrained, which is not surprising given the lack of information in the far-UV ($\lesssim2000$ \AA) region of the spectrum that would be sensitive to the inner disk radius.} For the other disc parameters, the outer radius $R_{out}$ and mass-transfer rate $\dot{m}$, we assumed log-uniform priors. We also imposed the condition that $R_{in}<R_{out}$.

The SED with the best fit is shown in Fig.~\ref{fig:SED_fit} (joint and marginal posterior distributions are shown in Appendix~\ref{sec:A1}). We found the outer radius of the disc to be well constrained $R_{out}=1.1^{+0.7}_{-0.3}\times10^{9}$ cm. The inclination of the system is unconstrained by our fits, but we can infer a 2$\sigma$ upper limit of $i < 80^{\circ}$, consistent with its X-ray spectral properties \citep{Degenaar:2016aa,eijnden:2017aa}. We obtain a $\dot{m}=7.2^{+0.2}_{-0.5}\times10^{-9}$~M$_{\odot}$ yr$^{-1}$, which suggest a high optical luminosity.  
This evident discrepancy between the $\dot{m}$ and that inferred from the low X-ray luminosity is discussed in Section~\ref{sec:discussion}.

We have included the best X-ray fitting components (\caps{bbody+gaussian+powerlaw}) to the simultaneous \textit{XMM-Newton} and \textit{NuStar} data presented in \citet{eijnden:2017aa} in Fig.~\ref{fig:SED_fit}. Our SED fitting suggests that the thermal component in the X-ray spectrum (\caps{bbody}) is highly unlikely to be from the disc, and more in line with thermal emission from the NS or boundary layer. This was previously proposed for this and other NSs accreting at low rates based on the inferred temperature and emitting radius of the thermal X-ray emission component \citep[e.g.][]{Armas-Padilla:2011aa,Wijnands:2015aa,Degenaar:2016aa},~but now for the first time demonstrated very clearly by studying the multi-wavelength SED.

\section{Discussion}\label{sec:discussion}
We have performed a multi-wavelength analysis of the AMXP and UCXB \igr. Our UV to NIR SED can be very well described by a standard accretion disc.  From our SED modelling, we measure an outer accretion disc radius of $2.2^{+0.9}_{-0.4} \times 10^{10}$ cm and a mass-transfer rate of $1.8^{+1.8}_{-0.5}\times10^{-10}$ M$_{\odot}$ yr$^{-1}$. Moreover, our SED modelling demonstrates that accretion disc spectrum does not extend into the soft X-rays. This implies that the thermal emission component that is seen in the X-ray spectrum of \igr\@ and other NS LMXBs accreting at low rates is likely from the surface of the NS, as was previously hypothesised based on X-ray spectral analysis \citep[e.g.][]{Armas-Padilla:2013aa,Degenaar:2016aa}. Furthermore, our low-resolution optical spectrum shows a blue-continuum consistent with an accretion disc, no emission lines of H, He or other elements are observed.  In the next sections we discuss the implications of our results.

\subsection{The system parameters of \igr}\label{sec:igr}
Theoretically, UCXBs can have different types of donor stars; an evolved (i.e. H-deficient) main sequence star, a He star, or a white dwarf (WD). Based on the measured orbital period and mass-accretion rate inferred from X-ray observations, \citet{Strohmayer:2017aa} favored a He donor for \igr. However, if significant mass loss occurs (i.e. if the mass-accretion rate is very different from the mass-transfer rate), the properties of \igr\@ can also be consistent with a CO WD donor. The latter scenario was proposed by \citet{eijnden:2017aa} based on an apparent overabundance of O inferred from high-resolution X-ray spectral fitting. 

Unfortunately, the lack of H and He features in our optical spectra do not directly constrain the donor type. Firstly, it does not fully rule out the presence of H and He in the accreted material; accretion disc spectral models suggest that He and H fractions up to $\sim$10\% might remain undetected in optical spectra \citep[][]{werner2006}. Moreover, depending on the amount of He that the donor burned before overflowing its Roche lobe, He donors may have relatively little He left and the accretion disc spectra of these systems may look similar to those harbouring CO donors \citep{Nelemans:2010mnras}. 

Understanding the composition of the matter accreted in UCXBs is of great value for understanding the thermonuclear burning of matter accreted onto the surface of NSs. Vice versa, the bursting behaviour can potentially be used to constrain the nature of the companion star \citep[e.g.][]{cumming2003}. Both bursts detected from \igr\@ were of unusually long duration and are indicative of the ignition of a thick layer of He. Such bursts are expected to occur for NSs that accrete He-rich material \citep[e.g.][]{intzand2005,cumming2006}. Another system that displays energetic and long X-ray bursts, 4U 0614+091, was suspected to harbour a CO white dwarf companion based on the lack of H and He features in its optical spectrum \citep[e.g.][]{nelemans2004,nelemans2006}. However, it was pointed out by  \citet{kuulkers2010} that it is difficult to produce its observed long He bursts if the donor is a CO WD with very little He. One possible solution would be to invoke additional physics, such as spallation \citep[e.g.][]{bildsten1992}. 

The case of 4U 0614+091 shows that the energetics of the thermonuclear X-ray bursts of \igr\@ do not unambiguously constrain the properties of its companion star. The recurrence time of thermonuclear X-ray burst may also contain information on the composition of the accreted matter. For instance, for the UCXB 4U 1820--30, the recurrence time between its thermonuclear X-ray bursts was used to investigate the evolutionary history of the binary \citep[][]{cumming2003}. However, the short thermonuclear X-ray bursts of 4U 1820--30 recur on a much shorter time scale (hours) than the long thermonuclear bursts that are observed for \igr\@ \citep[estimated to be months; e.g.][]{degenaar2010_rxh}. Indeed, only two bursts have been detected for \igr\@ and it is plausible that other bursts from these system have been missed. Therefore, we cannot put any meaningful constraints on the recurrence time of the bursts in \igr\@ and use that to investigate its evolutionary history. 

We note that the low X-ray luminosity of \igr\@ contrasts with the much higher mass-transfer rate from the donor implied by our SED modelling, as well as theoretical predictions for UCXBs~\citep[e.g.][]{Sengar:2017aa}. 
If the X-ray luminosity from the blackbody component is due to accretion onto the NS surface \citep{eijnden:2017aa}, $L_X\sim6\times10^{34}$ erg s$^{-1}$, we can obtain an estimate on the mass accretion rate. Assuming a totally efficient conversion of the accretion power into $L_X$, we find $\dot{m}_{NS}\sim3\times10^{-11}$ \myr.
Comparing this to the mass transfer rate of $1.8\times10^{-10}$ \myr\ inferred form our SED modelling, this would require that the system, in order for mass conservation to hold, ejects $\gtrsim90$\% of the in-falling material. 

One possible mechanism to reconcile this discrepancy is that the NS magnetosphere is truncating the inner accretion disc and acts as a magnetic propeller that drives an outflow \citep{Illarionov:1975aa}. High-resolution X-ray spectral analysis of \igr\ showed hints of a wind-like outflow in the form of oxygen-rich absorption and emission features~\citep{eijnden:2017aa}.~Several other LMXBs, including AMXPs, show (indirect) evidence for highly non-conservative mass transfer \citep[e.g.][]{disalvo2008,ponti2012,marino2017,tetarenko2018,ziolkowski2018}.

We conclude that our current multiwavelength data set, and the bursting behaviour of \igr, do not allow us to narrow down the nature of its companion star. However, this can potentially be achieved with additional observations. \citet{Nelemans:2010mnras} demonstrated that abundance ratios of N and C provide a strong diagnostic of the formation scenario of UCXBs, and even the mass of the progenitor of the donor star. Furthermore, N/O and O/C ratios can distinguish between a He and WD donor star. Our multiwavelength study and SED modelling show that \igr\@ is bright in the FUV.  Spectral FUV studies have been performed for a handful of (weakly extinct) LMXBs, and resulted in the detection of prominent emission lines of N, C, O, and He, giving valuable insight into the composition of the accreted matter and the evolutionary history of the binary \citep[e.g.][]{schulz2001,haswell2002,froning2011}. Obtaining a FUV spectrum of \igr\@ with {\it HST} is therefore expected to provide the opportunity to elucidate the composition of the accreted matter and the nature of the donor star.

\subsection{Multi-wavelength studies as a tool to find UCXBs}\label{sec:mw}
To date, only $\sim$2 dozen of UCXBs are known in our Galaxy \citep[e.g.][]{Nelemans:2010aa}. Among the LMXB population, the most promising candidates for identifying new (candidate) UCXBs are the systems that, like \igr,  persistently accrete at a very low rate \citep[e.g.][]{intzand2007,Degenaar:2016aa}; the so-called very-faint X-ray binaries (VFXBs). Therefore, we use our multi-wavelength data set on \igr\ to test indirect methods to constrain $P_{orb}$; verifying these methods for a system with known parameters could aid in observational campaigns of other VFXBs.
Such methods concern the relative contributions of the optical/infrared to X-ray fluxes \citep{van-Paradijs:1994aa,Revnivtsev:2012aa}. As the central X-ray source illuminates the accretion disc, a fraction of the incident energy is reprocessed, modifying the emitted SED. The intrinsic emission of the donor likely does not significantly contribute to the total luminosity.

\begin{figure}
	\includegraphics[trim=0.5cm 0.5cm 0.0cm 0.2cm, clip,width=\columnwidth,angle=270]{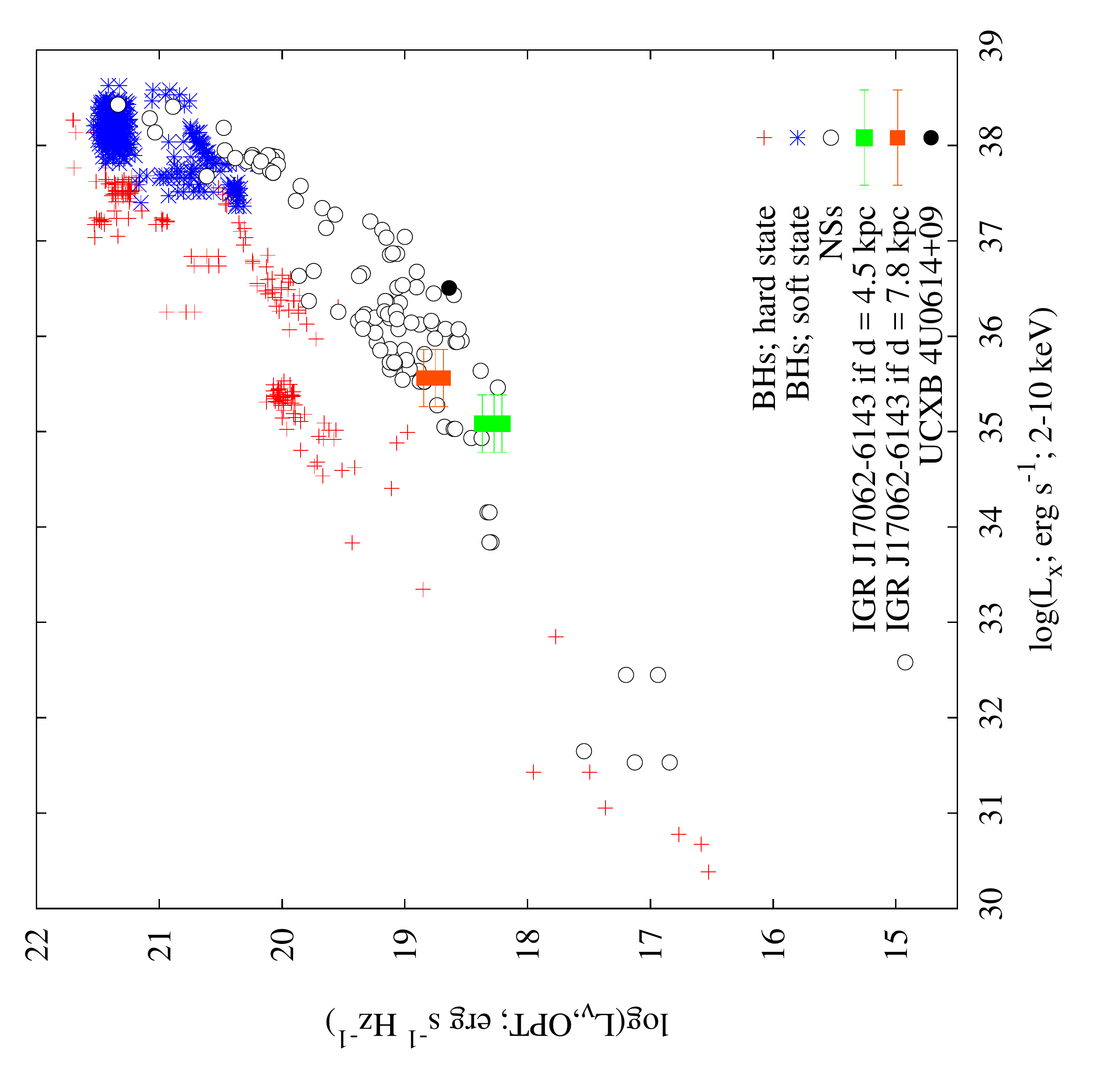}
    \caption{L$_x$ - L$_{\textrm{opt}}$ correlation for BH and NS. Data was taken from \citet{Russell:2006aa} and \citet{Russell:2007aa}. \igr\ (square) follows the NS track for both estimates of the distance to the system. We show the UCXB 4U06414+09 (filled circle) as a reference for a typical system.}
    \label{fig:opt_x}
\end{figure}

We have plotted the X-ray/optical luminosities of \igr\ against a large sample of BHs and NSs in Fig.~\ref{fig:opt_x}. As clearly shown, both classes of LMXBs have a different correlation, where BHs are a factor $\sim20$ brighter in optical than NS LMXBs \citep[][]{Russell:2006aa}. Taking into account both estimates of the distance, \igr\@ lies comfortably within the NS track. However, despite its position at the lower end of the NS track, we caution against any inference on the orbital period from the low optical luminosity alone \citep[e.g.][]{van-Paradijs:1994aa}. 
In particular, \igr\@'s luminosity is similar to other NS LMXBs with widely different orbital periods such as SAX J1808.4-3658 \citep[2.013 hr,][]{Chakrabarty:1998ab}, 4U 1608-52 \citep[12.89 hr,][]{Wachter:2002aa} and Aql X-1 \citep[18.95 hr,][]{Chevalier:1998aa}. 

On the other hand, the empirical $L_{K}-L_{X}$ relationship developed by \citet{Revnivtsev:2012aa} provides another framework to characterise persistent LMXBs, such as \igr. At NIR wavelengths, the system is assumed to be dominated by the reprocessing of the central X-ray flux in the accretion disc and the secondary star. The use of exclusive persistent NS-sources provides a more homogeneous data set (small variability and mass-dependence) to perform the calibration of this relationship.
This method describes the absolute magnitude in the $K$-band (in Vega system) as a function of the orbital period and X-ray luminosity,
\begin{equation}
M_K = (2.66\pm0.11)-2.5\log \Sigma_K\,,
\end{equation}
where
\begin{equation}
\Sigma_K=(L_X/L_{\rm Edd})^{0.29}(P_{\rm orb}[h])^{0.92}\,.
\end{equation}
In order to calculate the luminosity, we used a distance of $d=7.3\pm0.5$~kpc \citep{Keek:2016aa}, an extinction value of $E(B-V)=0.135\pm 0.5$, a correction on the extinction for the $K-$band  $A_K=0.11A_V$~\citep{Rieke:1985aa} and a fixed $L_X/L_{\rm Edd}=10^{-3}$. We find, for both NIR $K$-band measurements, an orbital period of $P_{orb}=0.64 \pm 0.20$ hr and $P_{orb} = 0.77 \pm 0.24$ hr\footnote{If we use a value for $d=5$ kpc \citep{Degenaar:2013aa}, we obtain shorter orbital periods $\sim0.2$ hr.}. The 1$\sigma$ uncertainties on the orbital periods were calculated via a Monte Carlo simulation.

\begin{figure}
	\includegraphics[trim=0.5cm 0.1cm 0.0cm 0.2cm, clip,width=\columnwidth]{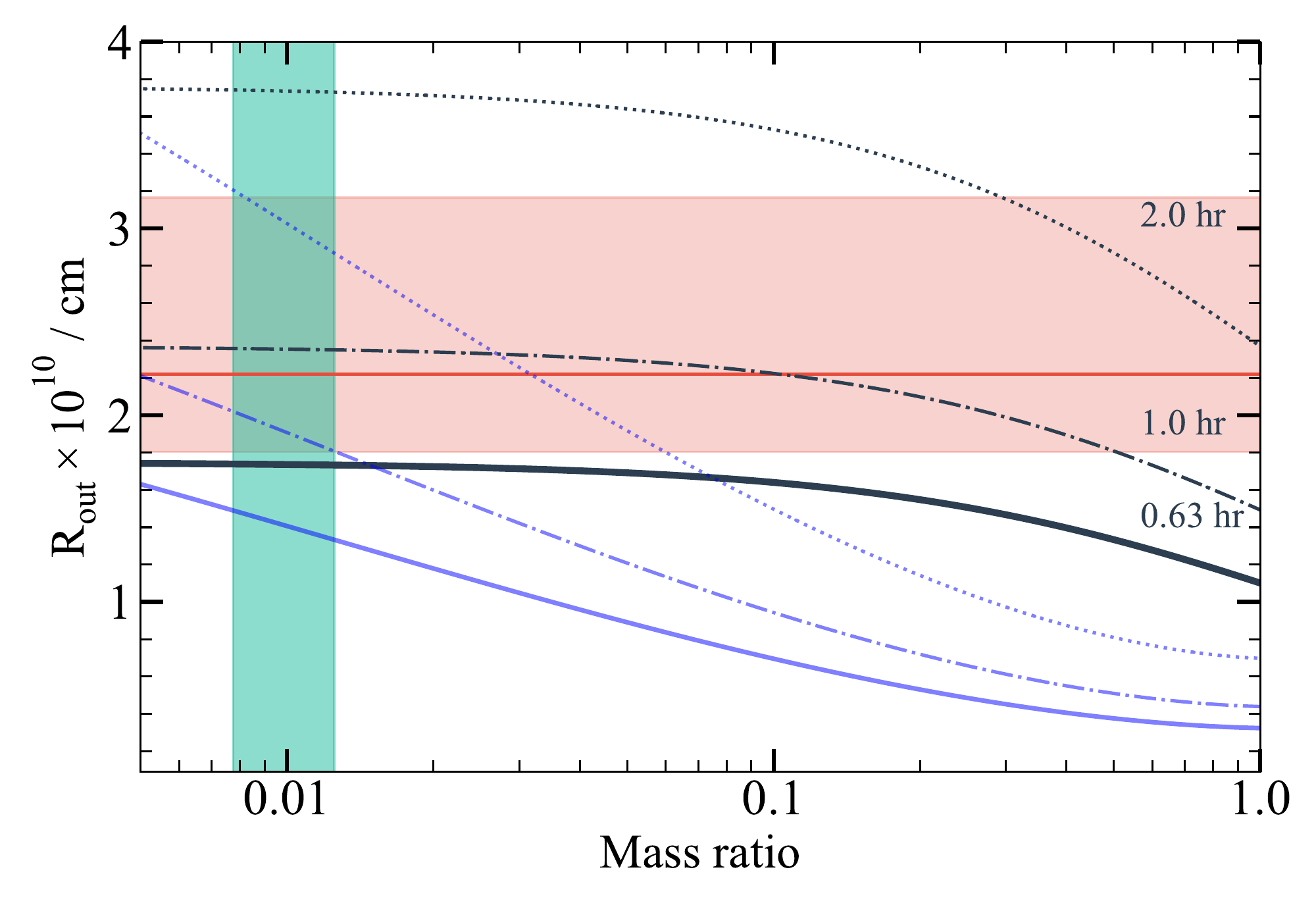}
    \caption{The tidal radius of an accretion disc as a function of mass ratio. We compare the $R_{out}$ estimate from the SED modelling (red line) to the predicted tidal (black lines) and circularisation (blue lines) radius for a given orbital period. The $1\sigma$ confidence levels of R$_{out}$ are shown as the red bands. The smallest orbital period represents the 38 minutes orbital period measured from X-ray pulsations and the green vertical band shows the range of mass ratios allowed from \citet{Strohmayer:2018aa}.}
    \label{fig:disc_size}
\end{figure}

Alternatively, we can constrain the orbital period by using the size of the accretion disc, measured in our SED modelling. The outer parts of the disc will be heavily affected by the tidal interaction of the companion thus regulating its maximum size \citep{Paczynski:1977aa}. This tidal radius, $r_{t}$, can be approximated by
\begin{equation}
\frac{r_{t}}{a} = \frac{0.6}{1+q}\,\,\,\,\,\,\,\,\,\,\,\,0.03<q<1,
\end{equation}
where $a$ is the orbital separation and $q=M_2/M_{NS}$ the mass ratio of the system. On the other hand, the minimum outer disc radius, $r_r$ is determined by the angular momentum of the particles after they exit the inner Lagrangian point. This can be approximated by the following equation \citep{Verbunt:1988aa}:
\begin{equation}
 \begin{split}
\frac{r_{r}}{a} = 0.0883 + 0.04858\log(q^{-1}) + \\ 
	0.11489\log^2(q^{-1}) - 0.020475\log^3(q^{-1}),
\end{split}
\end{equation} 
valid for $0.001\leq q<1$. We can find a family of solutions of $r_t$ and $r_r$ as a function of mass ratio given a fixed orbital period, as shown in Fig.~\ref{fig:disc_size}. Any allowed orbital period of the system should reside, for a given donor mass (assuming a canonical mass for the neutron star M$_{NS}=1.4$ M$_{\odot}$), between both curves. Despite the large uncertainties associated to the SED modelling, this methodology allows us to confidently exclude systems with P$_{orb}\gtrsim2$~hr (at more than 2$\sigma$) and highly suggestive of a system with P$_{orb}\lesssim1$~hr since the estimates of the disc are smaller than the minimum radius $r_r$. 

We conclude that the multi-epoch UV, optical and NIR photometry allows us to model the accretion disc and retrieve physical parameters of the binary system, such as the outer disc radius and mass transfer, that point to a system with an orbital period of $\lesssim1$~hr. In addition, we employed empirical relations to estimate an orbital period between 0.3-0.5 hr for \igr. Both techniques thus yield results that are consistent with the directly measured orbital period of 37.96 min from timing the X-ray pulsations. This demonstrates that multi-wavelength observations can effectively be used to search for UCXBs among VFXBs.

\section*{Acknowledgements}
We thank the referees for valuable comments and suggestions that significantly improved this work. JVHS, ND, and JvE are supported by a Vidi grant awarded to ND by the Netherlands Organization for Scientific Research (NWO). JVHS acknowledges partial support from ``NewCompStar", COST Action MP1304 and thanks the IoA, Cambridge for their hospitality during the exchange visit. DA acknowledges support from the Royal Society. RW is supported by an NWO Top grant, module 1. VC is supported by a grant awarded by the Consejo Nacional de Investigaciones Cient\'ificas y T\'ecnicas (CONICET), Argentina. We thank Andy Monson for developing and supporting FSRED and for help installing that package. We are grateful to Neil Gehrels and the Swift duty scientists for making ToO observations of \igr\ possible. We also thank German Gimeno and the Gemini duty scientists for the observation of \igr\ under a FT observing program. The Faulkes Telescopes are maintained and operated by the Las Cumbres Observatory (LCO). We acknowledge the use of public data from the Swift data archive. This research made use of \caps{astropy}, a community-developed core Python package for Astronomy \citep{Astropy-Collaboration:2013aa}, \caps{matplotlib} \citep{Hunter:2007aa} and \caps{aplpy} \citep{Robitaille:2012aa}.




\bibliographystyle{mnras}
\bibliography{bibliography.bib} 




\appendix

\section{Swift photometry}\label{sec:A2}
We present in Table~\ref{tab:flux_swift} all the individual measurements obtained with \textit{Swift/UVOT} used in the SED modelling in AB mag with the X-ray flux associated for each pointing. Details on the data reduction can be found in Section~\ref{sec:uvot} for UVOT and Section~\ref{sec:xrt} for XRT.
\begin{table*}
	\centering
	\caption{UV, optical and X-ray photometry from \textit{Swift}. Since the \textit{Swift} X-ray flux measurements were performed for each individual snapshot, there are UVOT entries with repeated $F_x$ values.}
	\label{tab:flux_swift}
\begin{tabular}{cccccc}
\hline
Filter & MJD & Exposure Time & Magnitude & $F_{\nu}$ & $F_x$ [2-10 keV]  \\ 
 &  & s &  UVOT System &  mJy  &  $\times 10^{-10}$ erg s$^{-1}$ cm$^{-2}$  \\ 

\hline
UVW2 & 54587.45660 &   277.0 &   19.97 $\pm$    0.12 &   0.037 $\pm$   0.004 &   1.478 $\pm$   0.148 \\
UVW2 & 54588.19469 &  1061.9 &   20.11 $\pm$    0.08 &   0.033 $\pm$   0.002 &   1.268 $\pm$   0.127 \\
UVW2 & 54588.05937 &   283.2 &   20.18 $\pm$    0.13 &   0.031 $\pm$   0.004 &   1.268 $\pm$   0.127 \\
UVW2 & 54592.43834 &   790.6 &   20.15 $\pm$    0.08 &   0.032 $\pm$   0.002 &   2.325 $\pm$   0.232 \\
UVW2 & 54592.33448 &   224.1 &   20.07 $\pm$    0.14 &   0.034 $\pm$   0.004 &   2.325 $\pm$   0.232 \\
UVW2 & 54593.74933 &  1184.3 &   20.13 $\pm$    0.07 &   0.032 $\pm$   0.002 &   1.991 $\pm$   0.199 \\
UVW2 & 54593.61518 &   460.4 &   20.15 $\pm$    0.11 &   0.032 $\pm$   0.003 &   1.991 $\pm$   0.199 \\
UVW2 & 54594.61436 &  2391.6 &   20.08 $\pm$    0.06 &   0.034 $\pm$   0.002 &   1.991 $\pm$   0.199 \\
UVW2 & 54594.27713 &    66.7 &   20.08 $\pm$    0.24 &   0.034 $\pm$   0.007 &   1.991 $\pm$   0.199 \\
UVW2 & 54595.28439 &  3163.7 &   20.08 $\pm$    0.05 &   0.034 $\pm$   0.002 &   1.934 $\pm$   0.193 \\
UVW2 & 54595.01637 &   558.8 &   20.15 $\pm$    0.10 &   0.032 $\pm$   0.003 &   1.157 $\pm$   0.116 \\
UVW2 & 54597.42468 &  4067.9 &   20.11 $\pm$    0.05 &   0.033 $\pm$   0.002 &   1.576 $\pm$   0.158 \\
UVW2 & 54597.02051 &   627.1 &   20.18 $\pm$    0.09 &   0.031 $\pm$   0.003 &   1.576 $\pm$   0.158 \\
UVW2 & 54661.53944 &  6809.6 &   20.03 $\pm$    0.05 &   0.035 $\pm$   0.001 &   1.598 $\pm$   0.160 \\
UVW2 & 54661.33972 &  1021.9 &   20.08 $\pm$    0.07 &   0.034 $\pm$   0.002 &   1.598 $\pm$   0.160 \\
\hline
UVM2 & 57624.60002 &   805.1 &   20.72 $\pm$    0.16 &   0.019 $\pm$   0.003 &   0.546 $\pm$   0.055 \\
UVM2 & 57630.18016 &   992.4 &   20.90 $\pm$    0.16 &   0.016 $\pm$   0.002 &   0.306 $\pm$   0.031 \\
UVM2 & 57639.42057 &  1088.9 &   20.40 $\pm$    0.11 &   0.025 $\pm$   0.003 &   0.358 $\pm$   0.036 \\
UVM2 & 57646.85150 &   551.5 &   20.97 $\pm$    0.23 &   0.015 $\pm$   0.003 &   0.323 $\pm$   0.032 \\
UVM2 & 54587.78830 &    82.7 &   20.46 $\pm$    0.37 &   0.024 $\pm$   0.008 &   1.268 $\pm$   0.127 \\
UVM2 & 54588.06322 &   213.8 &   20.12 $\pm$    0.19 &   0.033 $\pm$   0.006 &   1.268 $\pm$   0.127 \\
UVM2 & 54592.33732 &   130.4 &   19.97 $\pm$    0.22 &   0.037 $\pm$   0.008 &   2.325 $\pm$   0.232 \\
UVM2 & 54593.62108 &   294.3 &   19.80 $\pm$    0.14 &   0.044 $\pm$   0.006 &   1.991 $\pm$   0.199 \\
UVM2 & 54594.27809 &    47.5 &   20.91 $\pm$    0.64 &   0.016 $\pm$   0.009 &   1.991 $\pm$   0.199 \\
UVM2 & 54595.02385 &   414.6 &   20.16 $\pm$    0.14 &   0.031 $\pm$   0.004 &   1.157 $\pm$   0.116 \\
UVM2 & 54598.03805 &   833.0 &   19.99 $\pm$    0.09 &   0.037 $\pm$   0.003 &   1.576 $\pm$   0.158 \\
\hline
UVW1 & 54587.45231 &   141.5 &   20.04 $\pm$    0.19 &   0.034 $\pm$   0.006 &   1.478 $\pm$   0.148 \\
UVW1 & 54588.05504 &   141.5 &   20.11 $\pm$    0.20 &   0.032 $\pm$   0.006 &   1.268 $\pm$   0.127 \\
UVW1 & 54590.10843 &   437.2 &   19.93 $\pm$    0.12 &   0.038 $\pm$   0.004 &   1.032 $\pm$   0.103 \\
UVW1 & 54590.00866 &   121.8 &   20.44 $\pm$    0.31 &   0.024 $\pm$   0.007 &   1.032 $\pm$   0.103 \\
UVW1 & 54592.14843 &   130.5 &   20.08 $\pm$    0.22 &   0.033 $\pm$   0.007 &   2.325 $\pm$   0.232 \\
UVW1 & 54593.60824 &   230.1 &   19.86 $\pm$    0.14 &   0.041 $\pm$   0.005 &   1.991 $\pm$   0.199 \\
UVW1 & 54594.27599 &    33.2 &   19.93 $\pm$    0.36 &   0.038 $\pm$   0.013 &   1.991 $\pm$   0.199 \\
UVW1 & 54595.00798 &   279.3 &   19.83 $\pm$    0.12 &   0.042 $\pm$   0.005 &   1.157 $\pm$   0.116 \\
\hline
U & 54587.45361 &    70.6 &   19.87 $\pm$    0.26 &   0.042 $\pm$   0.010 &   1.478 $\pm$   0.148 \\
U & 54588.05633 &    70.6 &   19.73 $\pm$    0.24 &   0.047 $\pm$   0.010 &   1.268 $\pm$   0.127 \\
U & 54590.00952 &    16.4 &   20.55 $\pm$    1.57 &   0.022 $\pm$   0.032 &   1.032 $\pm$   0.103 \\
U & 54592.33206 &    55.9 &   19.45 $\pm$    0.22 &   0.061 $\pm$   0.012 &   2.325 $\pm$   0.232 \\
U & 54593.61033 &   114.9 &   19.56 $\pm$    0.17 &   0.055 $\pm$   0.009 &   1.991 $\pm$   0.199 \\
U & 54594.27633 &    16.5 &   19.10 $\pm$    0.31 &   0.084 $\pm$   0.024 &   1.991 $\pm$   0.199 \\
U & 54595.01050 &   139.5 &   19.87 $\pm$    0.19 &   0.042 $\pm$   0.007 &   1.157 $\pm$   0.116 \\
U & 54600.41665 &  3073.0 &   19.68 $\pm$    0.05 &   0.050 $\pm$   0.002 &   1.548 $\pm$   0.155 \\
U & 54600.18262 &   339.6 &   19.74 $\pm$    0.15 &   0.047 $\pm$   0.007 &   1.675 $\pm$   0.167 \\
U & 54660.40377 &   993.9 &   19.70 $\pm$    0.07 &   0.049 $\pm$   0.003 &   1.598 $\pm$   0.160 \\
U & 57644.49897 &  1932.3 &   20.38 $\pm$    0.10 &   0.026 $\pm$   0.002 &   0.435 $\pm$   0.044 \\
U & 57644.46425 &   982.6 &   20.57 $\pm$    0.17 &   0.022 $\pm$   0.003 &   0.435 $\pm$   0.044 \\
\hline
B & 54587.45449 &    70.6 &   20.04 $\pm$    0.56 &   0.035 $\pm$   0.018 &   1.478 $\pm$   0.148 \\
B & 54588.05722 &    70.6 &   20.73 $\pm$    1.02 &   0.018 $\pm$   0.017 &   1.268 $\pm$   0.127 \\
B & 54592.33277 &    55.9 &   19.36 $\pm$    0.36 &   0.065 $\pm$   0.021 &   2.325 $\pm$   0.232 \\
B & 54593.61173 &   114.9 &   20.15 $\pm$    0.49 &   0.031 $\pm$   0.014 &   1.991 $\pm$   0.199 \\
B & 54594.27658 &    16.5 &   21.72 $\pm$    5.05 &   0.007 $\pm$   0.034 &   1.991 $\pm$   0.199 \\
B & 54595.01220 &   139.5 &   19.80 $\pm$    0.32 &   0.043 $\pm$   0.013 &   1.157 $\pm$   0.116 \\
\hline
	\end{tabular}
\end{table*}

\section{SED Posterior distributions}\label{sec:A1}

\begin{figure*}
	\includegraphics[width=15cm]{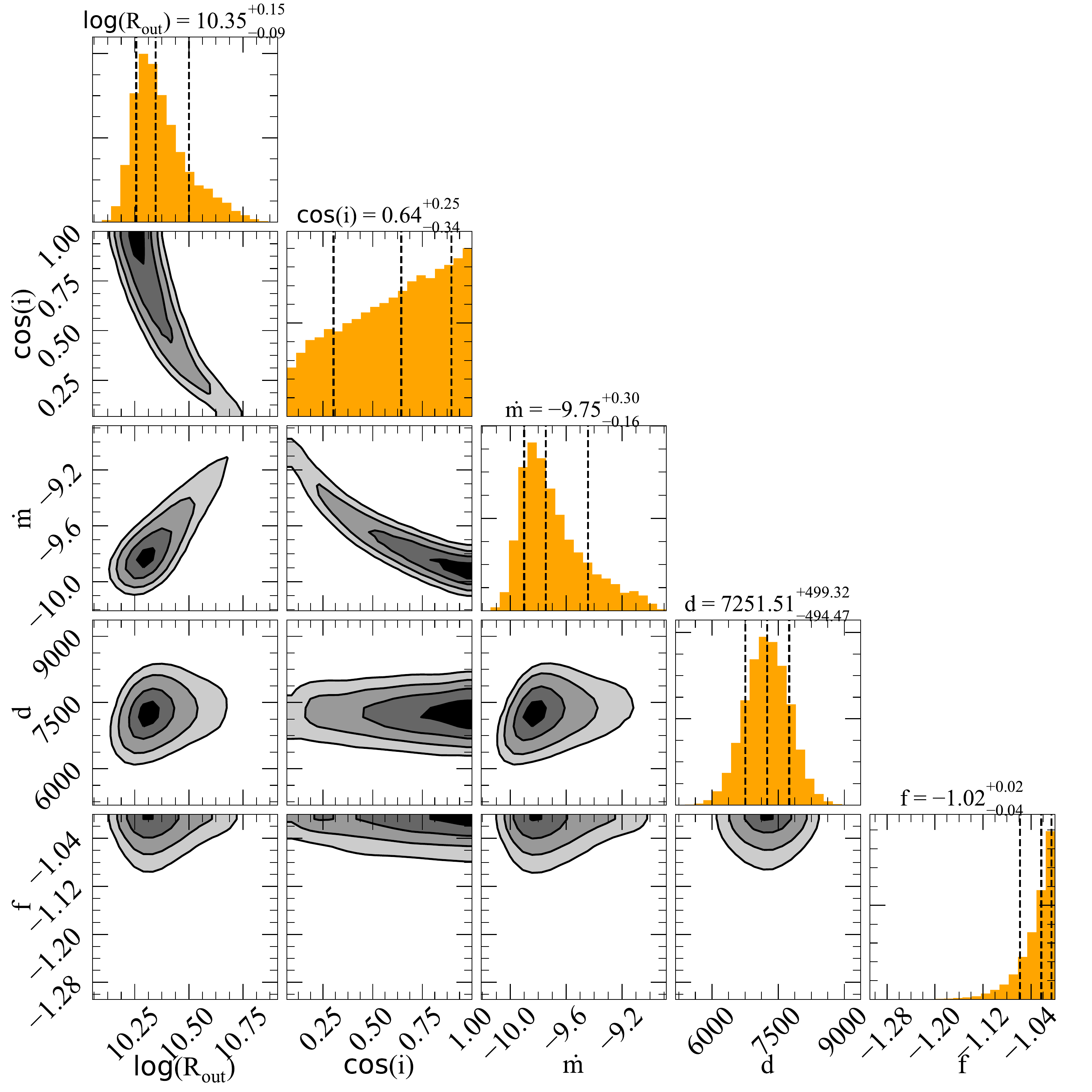}
    \caption{Posterior probability distributions for the accretion disc parameters. Colour scale contours show the joint probability for every combination of parameters. Contours represent the 0.5$\sigma$, 1$\sigma$ , 2$\sigma$ and 3$\sigma$ levels. Marginal posterior distributions are shown as histograms with the median and 1$\sigma$ marked as dashed lines. }
    \label{fig:posteriors}
\end{figure*}

We present the joint and marginal posterior distributions of the accretion disc parameters obtained in the SED fit in Fig.~\ref{fig:posteriors} (see Sec.~\ref{sec:sed}). We used \caps{corner.py} \citep{corner} to visualise the MCMC chains.

\bsp	
\label{lastpage}
\end{document}